\documentclass[onecolumn]{emulateapj}
\slugcomment{}

\shorttitle{Simulations of Flares} \shortauthors{Tang \& Smith}

\begin{document}

\title{GEANT4 Simulations of Gamma-Ray Emission from Accelerated Particles in Solar Flares}

\author{ShiChao Tang\altaffilmark{1} and David M. Smith\altaffilmark{2,3}}
\altaffiltext{1}{Department of Physics \&
  Center for Astrophysics, Tsinghua University, Beijing, 10084, China}
\altaffiltext{2}{Physics Department and Santa Cruz Institute for
Particle Physics, University of California, Santa Cruz, Santa Cruz,
CA 95064, USA}
\altaffiltext{3}{Space Sciences Laboratory, University of California,
Berkeley, Berkeley, CA 94720, USA}
\begin{abstract} 

Gamma-ray spectroscopy provides diagnostics of particle acceleration in
solar flares, but care must be taken when interpreting the spectra due to
effects of the angular distribution of the accelerated particles (such
as relativistic beaming) and Compton reprocessing of the radiation in
the solar atmosphere.  In this paper, we use the GEANT4 Monte Carlo
package to simulate the interactions of accelerated electrons and
protons and study these effects on the gamma-rays resulting from
electron bremsstrahlung and pion decay.  We consider the ratio of the
511~keV annihilation-line flux to the continuum at 200~keV and in the
energy band just above the nuclear de-excitation lines (8--15~MeV) as
a diagnostic of the accelerated particles and a point of comparison
with data from the X17 flare of 2003 October 28.  We also find that
pion secondaries from accelerated protons produce a positron
annihilation line component at a depth of $\sim$ 10 g cm$^{-2}$, and
that the subsequent Compton scattering of the 511~keV photons produces
a continuum that can mimic the spectrum expected from the 3$\gamma$
decay of orthopositronium.

\end{abstract}

\keywords{gamma rays --- radiation mechanisms: nonthermal ---Sun: flares --- Sun:
  X-rays }

\section{Introduction}

In solar flares, electrons and ions are accelerated to non-thermal
energies. When these particles interact with the ambient medium, they
can produce photons with energies up to the gamma-ray range. The
electrons produce continuum emission via the bremsstrahlung process,
while the ions (protons and heavier nuclei) can produce excited and radioactive nuclei which,
through de-excitation or decay, make emission lines usually $\lesssim$
7~MeV. Ions with energy above $\sim$ 200~MeV can produce pions by
interacting with ambient nuclei.  These pions then
produce gamma-ray continuum via $\pi^0\rightarrow 2\gamma$ or
$\pi^{\pm} \rightarrow \mu^{\pm} \rightarrow e^{\pm}\rightarrow
\gamma_{brem}$. There is also 511~keV line emission from annihilation
of positrons created by the decay of $\beta^{+}$-emitting radioactive
nuclei or $\pi^{+}$. Whether radioactive nuclei or $\pi^{+}$
contribute more to the positron population depends on the hardness of
the injected ions \citep{murphy84}.  Positrons may also come from the
$e^{-}e^{+}$ pair production process of the gamma-ray continuum.

These continua and lines provide information on particle acceleration
in solar flares, but we can only observe those photons that reach us.
Since the accelerated electrons are relativistic, the angular
distribution of bremsstrahlung will tend to follow that of the
original electrons, so electrons beamed downward along the magnetic
field will put most of their radiation into the Sun.  Photons created
deep in the solar atmosphere by any process are less likely to escape
than those created in the corona or chromosphere.  These effects will
change the relative luminosity of different spectral components, but
the location and directionality of the photon production processes
will also change the spectral shape of each as well.  Bremsstrahlung
intrinsically creates different spectra in different directions, with
the hardest spectrum in the beam direction.  Compton scattering can
also affect the observed spectra \citep[e.g.]{kotoku07} by scattering
photons to lower energy. The importance of this process depends on
both the depth where the original photons are produced and their
direction relative to the line of sight.  Accurate simulations of the
location, beaming, reprocessing, and absorption of flare photons are
therefore just as important to interpreting spectra as modeling of the
original radiation mechanism.

Examples of the importance of directionality and location in
interpreting observed bremsstrahlung spectra are seen in recent work
by \citet{krucker08} and \citet{kontar06}.  The picture of electrons
accelerated directly down field lines into the deep solar atmosphere
is seldom found to agree with observations.  Strong scattering, as
from interactions with magnetohydrodynamic waves, can make the
distribution function evolve into one that is more isotropic than when
the electrons were injected into the magnetic loop.  \citet{bret09}
compared different kinds of instabilities that may cause this
anisotropy decrease. The evolution speed of the distribution function
is sensitive to the loop magnetic field as described by
\citet{karlicky09}.  Magnetic mirroring can produce a ``pancake''
distribution moving mostly parallel to the solar surface
\citep{dermer86}.  \citet{krucker08} proposed a scenario in which the
highest energy bremsstrahlung in flares is produced in the magnetic
loop top, because they found that the coronal source is harder and
becomes dominant above 500~keV in the 2005 January 20 flare.  In this
picture, the angular distribution of gamma-ray producing electrons is
also isotropic because they are trapped by strong scattering.
\citet{kontar06} concluded that the angular distribution of electrons
producing hard X-rays at flare footpoints is isotropic by including
the Compton-scattered X-ray ``albedo'' surrounding the footpoints in
their spectral analysis.

In the present work, we perform Monte Carlo simulations with the
toolkit GEANT4 \citep{agostinelli03} to illustrate the effects of
beaming and reprocessing on observable gamma-ray components from
flare-accelerated electrons and protons.  We focus on electron
bremsstrahlung and the secondary radiation from pion production by
protons, since we believe that GEANT4 addresses these components more
accurately than it does the nuclear de-excitation lines that dominate
between these energy ranges.  In particular, we study the
positron-annihilation line and the production of continuum radiation
from 8--15~MeV, a range bracketed at the bottom by the energy at which
de-excitation lines first become negligible and at the top by the
maximum energy observed by the {\it Reuven Ramaty High-Energy Solar
Spectroscopic Imager (RHESSI)} \citep{lin02}, which we will use to
compare our simulations to a flare observation.  If the 8--15~MeV
continuum is taken to be bremsstrahlung from flare-accelerated
electrons, we show that there is a strong constraint on the angular
distribution of the electrons, an effect discussed also by
\citet{kotoku07}.

The positron-annihilation line is always isotropic, because the
positrons mostly slow down to thermal speed before they annihilate.
Thus, a comparison of the annihilation line and the 8--15~MeV
continuum gives us further information on the electron angular
distribution, again under the assumption that both components are due
to bremsstrahlung and its reprocessing.  We will also discuss the
other sources for these energy bands: for the line, $\beta+$ decay and
positrons from $\pi+$, and for the continuum, bremsstrahlung from
electrons and positrons from the decay of charged pions and
reprocessing of gamma-rays from the decay of neutral pions in the
solar atmosphere and in the instrument.  Positrons can annihilate
through the 3$\gamma$ orthopositronium channel.  The resulting
continuum, while it has a characteristic shape, can be mistaken for
Comptonization of the 511~keV line \citep{share04} and can greatly
affect the estimation of the ratio between the annihilation line and
other spectral components.

For an example both of the capabilities of these simulations and of
instrumental effects, we use an observation of the large X-class flare
of 2003 October 28 with {\it RHESSI} and compare the time
  integrated spectrum with our simulations.

The models and method are shown in detail in \S\ref{sec:model}. In
\S\ref{sec:simu} we show the simulation results and compare them with
the 2003 October 28 flare, and \S\ref{sec:dis} provides the summary
and discussion.

\section{Model and method}
\label{sec:model}

\subsection{The GEANT4 tool kit}
\label{sec:geant}

We used the Monte Carlo simulation package GEANT4
\citep{agostinelli03}, which is widely used in experimental
high-energy physics for simulating the passage of particles through
matter. The physics processes offered cover all the electromagnetic
and hadronic processes we are interested in.  GEANT4 treats individual
simulated particles one at a time rather than distributions of
particles, and carries them through a mass model of the universe
defined by geometrical boundaries between materials rather than a
grid.  When a particle is ``injected'', GEANT4 calculates the mean
free path of all the discrete physics processes implemented,
calculates a random distance associated with each process, and chooses
that with the shortest distance to be implemented (unless the distance
to the nearest material boundary is closer, in which case the particle
is taken to the boundary).  It then determines all the physics
properties of the particle after the chosen process (including its new
position) taking account of the continual physics processes (such as
energy loss by electrons) that happen within this step. As this goes
on, it provides a ``track'' of the particle until it comes to rest,
leaves the volume, or reaches a low-energy threshold.  Daughter
particles, when created, are tracked immediately, with the parent
particle put aside to be followed after the daughter particle is
finished.  Cascades can thus be followed deeply, restricted only by
computer memory.  In our simulations, we inject electrons or protons
into a model of the solar atmosphere and track their interactions with
the ambient material, recording the angular and energy distribution of
photons leaving the Sun.

\citet{kotoku07} used GEANT4 to simulate electron bremsstrahlung in
solar flares, and included the effect of Compton scattering on the
emerging bremsstrahlung continuum.  In this work, we also quantify the
positron-annihilation line resulting from bremsstrahlung photons
pair-producing in the Sun, and simulate the gamma-ray emissions
originating from accelerated protons as well, considering the
continuum and annihilation-line photons resulting from pion creation.

\subsubsection{Electromagnetic processes}
\label{sec:em}

GEANT4 allows the user to select the particular physics processes to
be used, including alternate versions of some processes.  The
following electromagnetic processes are implemented in our
simulations: bremsstrahlung, ionization and Coulomb scattering for
electrons and positrons, and pair-production, Compton
scattering, and photoelectric absorption for photons (the latter is
not significant at the energies of interest). Because we focus
on very high energies, we treat the chromosphere as a cold target even
though its temperature could reach as high as tens of keV when bombarded by
flare-accelerated particles.

Annihilation is also implemented for positrons, but without the
formation of positronium, so the annihilation always produces two
gamma-ray photons of 511~keV.  Some fraction of positrons in the real
solar atmosphere may form parapositronium (which also decays to two
511~keV photons) and orthopositronium (which decays to three
continuum photons with maximum energy 511~keV), but the
orthopositronium continuum can be quenched by collisions at high
densities; see \citet{murphy05} for extensive recent calculations.  We
will return to this problem in section \ref{sec:oct28} while comparing
our simulations with a flare observation.

The processes mentioned above, except for Coulomb scattering, are
implemented through the Penelope physics package \citep{salvat06},
which is valid above $\sim$ 250~eV.  Bremsstrahlung in Penelope
includes electron-electron bremsstrahlung, which can be significant
above several hundred keV in flares \citep{kontar07}, using cross
sections calculated after \citet{seltzer85}.  Coulomb scattering needs
special consideration because its mean free path is much smaller than
that of any other process and a full treatment would take too much
time. GEANT4 simulates the effect of multiple Coulomb interactions
after a given step as a statistical expectation, instead of treating
them one by one.

\subsubsection{Hadronic processes}
\label{sec:hadron}

\begin{figure}[htbp]
\centerline{\includegraphics[width=0.6\textwidth]{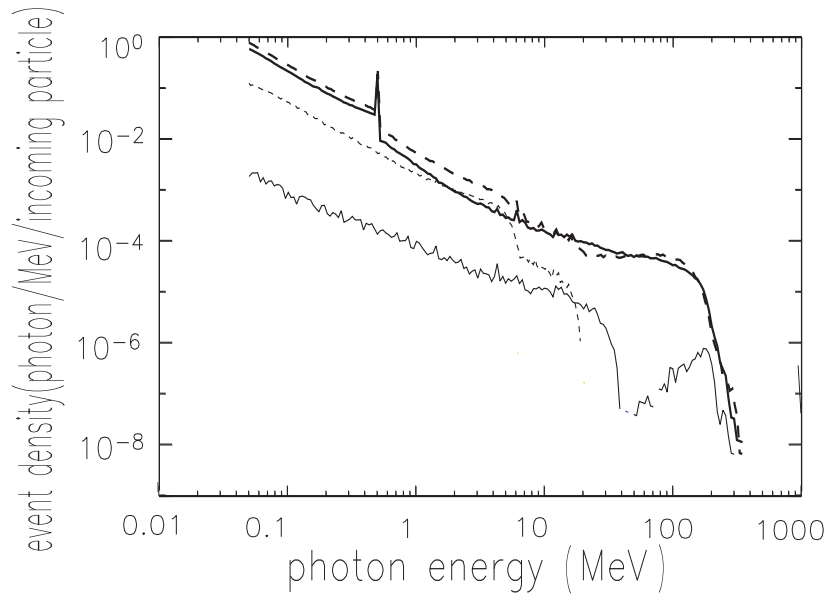}}
\caption[] {A comparison between the
  nuclear de-excitation spectra generated
  with two different physics modules in GEANT4. The
  details of the simulation parameters are in the text. 
  The thick and thin solid lines represent the total spectrum and the
  contribution of proton inelastic processes respectively, from the module
  HadronPhysicsQGSP\_BERT\_HP. The thick and thin dashed lines are
  from the module HadronPhysicsQGSP\_BIC\_HP.}\label{fig:modcompare}
\end{figure}

\citet{chin09} suggested that because the different GEANT4 models for
proton inelastic collisions are not consistent with each other in the
energy range of tens of MeV, these models are not ready for problem
solving; our initial simulations using these packages confirm this
conclusion for our application. We found that the proton inelastic
process as simulated in the HadronPhysicsQGSP\_BERT\_HP physics module
produces a continuum-like photon spectrum extending to $\sim$ 20 MeV.
This conflicts with most observations of gamma-ray flares
\citep[e.g.]{share95}, in which this component falls off dramatically
at 8~MeV, above the complex of lines from the de-excitation of
nitrogen, carbon and oxygen.  The observed spectrum is also more
dominated by individual lines and looks less like a continuum.  Another
package, HadronPhysicsQGSP\_BIC\_HP, shows a more realistic cutoff
around 8~MeV but still produces only a continuum-like shape and not
the observed complex of de-excitation lines.  In Figure \ref{fig:modcompare}
we show the difference between these two physics modules for the
proton inelastic scattering component and the entire proton-derived
flux, including the high-energy component from pion production.  For
this comparison we injected mono-energetic protons of 1 GeV downward
into the model Sun described in section \ref{sec:solarmod} and
recorded all the photons coming out at an angle $\beta$ from the solar
normal such that $\cos(\beta)>0.8$.

\begin{figure}[htbp]
\centerline{\includegraphics[width=0.6\textwidth]{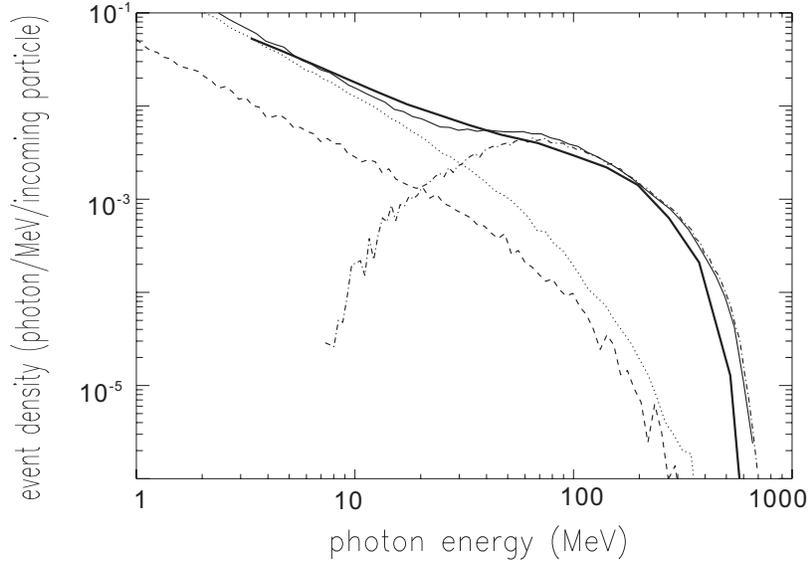}}
\caption[] {A comparison between the pion-decay spectra
  generated with the HadronPhysicsQGSP\_BERT\_HP module of GEANT4
  and the code developed at NRL by
  R. Murphy et al.; the  HadronPhysicsQGSP\_BIC\_HP module gives 
  similar results.  \textit{Dash dot:} $\pi^0$ decay.
  \textit{Dotted:} bremsstrahlung of $e^{\pm}$ from the decay of
  $\pi^{\pm}$.  \textit{Dashed:} positron annihilation.
  \textit{Thin solid:} the total spectrum from GEANT4. 
  \textit{Thick solid:} the total spectrum from the NRL code.}\label{fig:pioncompare}
\end{figure}

Pion production and decay, on the other hand, are simpler processes
and there is agreement among multiple models for their simulation.  As
can be seen in Figure \ref{fig:modcompare}, the two GEANT4 physics
modules agree well as to this photon component (dominant above
10~MeV).  We also compared these results from GEANT4 with the output
of the code developed by Reuven Ramaty, Ronald Murphy, and others,
which has been successfully applied to large flares \citep{murphy87}.
For this comparison, we put mono-energetic protons of 1~GeV into a
atmosphere of hydrogen and helium with He/H = 0.1, and recorded all
the photons produced. As shown in Figure \ref{fig:pioncompare}, the
results are consistent with GEANT4;  in the
energy range 8--15~MeV, the difference is less than 20\%.  Since we
believe we can therefore trust GEANT4 for pion processes as well as
electromagnetic cascades, we will ignore the inelastic process for
this work and concentrate on the 8--15~MeV range, which should be
dominated by bremsstrahlung both from primary electrons and from secondary
electrons and positrons from pion decay.

\subsection{The solar model}
\label{sec:solarmod}

In the present simulation, we treat the solar atmosphere as parallel
layers. This is a reasonable approximation because the size of the
emission region is always much smaller than the solar radius.

We use the analytical approximation of \citet{kotoku07} to the
Harvard-Smithsonian reference atmosphere \citep{gingerich71} as the
solar mass-density profile, which is
\begin{eqnarray}
  \label{eq:density}
  \rho(z)=3.19\times10^{-7}\exp(-\frac{z}{h})~~\mathrm{g~cm^{-3}}.
\end{eqnarray}
Here $z$ is the height measured from photosphere and $h$ is the scale
height, which is $\sim 400$ km~for $z<$0 and $\sim 110$~km for $z>$0.

The specific vertical structure model will not affect the simulation
result, however. All the processes we care about except decays depend
only on the column density along the path. Because our model solar
atmosphere is made up of parallel slabs, we can always transform the
parameter $z$ into column depth without changing our results.  The
only exception is decay processes, which depend on time. However, the
longest lifetime we need to consider is that of $\pi^{\pm}$, which is
2.6$\times 10^{-8}$s.  During this short time, the pion travels less
than 1~km even if it has an energy of up to 1$\times10^{4}$~MeV, which
is much shorter than the length scale of the system. So the decay
always happens approximately where the short lived particles are
produced and the result will not change with different vertical
structures. Another consideration is that when the chromosphere is
bombarded by the flare accelerated particles, it will evaporate and
fill up the magnetic loop. This will change the vertical structure. If
the particles were to interact high in a narrow column, the pattern of
photon escape as a function of solar normal angle would be very
different, with tangential escape much easier.  However,
\citet{aschwanden97} find that the density of evaporation upflow is
around 10$^{10}$ cm$^{-3}$. Take the longest flare loop with a length
of around 10$^{11}$ cm, the column density change will be less than
0.01 gcm$^{-3}$. As we will show in \S \ref{sec:simu}, all the
processes we are interested in happen at a column density greater that
1 gcm$^{-3}$. Therefore, evaporation should not strongly influence our
results.

The elemental abundance of our model Sun is taken from
\citet{grevesse07}, and we assume that the abundances of the
photosphere and corona are the same.  In future work we will implement
more realistic photospheric and coronal abundances, including the
enhancement of low first ionization potential (FIP) elements in the
corona.  \citet{kotoku07} used a pure hydrogen atmosphere, which
underestimates bremsstrahlung efficiency, since that rises as
approximately the square of atomic number $Z$, and also
underestimates pair production, the cross-section for which increases
even dramatically with $Z$.  We find that 10~MeV photons, for
example, produce 12\% more positrons in a realistic atmosphere than
in a hydrogen atmosphere.

\section{Results}
\label{sec:result}

\subsection{Simulations}
\label{sec:simu}

We inject electrons and protons with different energies and angular
distributions into our model Sun and track them as well as their
secondaries. The tracking stops if the particles leave the Sun or if
their energy falls below 50~keV. However, because positrons seldom
annihilate until they thermalize, the 50~keV cutoff does not apply to
them; we track them until they annihilate or leave the Sun. We also
track pions to their decay. 

\subsubsection{Interactions of Accelerated Electrons}
\label{sec:ele}

Table \ref{tab:e} shows all the different models of injected electrons
used in the simulations.  For the downward-beamed and
downward-isotropic distributions, the electrons are initialized just
above the model solar atmosphere.  For the isotropic and pancake
distributions, they are injected at an integrated column depth of $8
\times 10^{-5}$ g cm$^{-2}$.  The results are not sensitive to this
parameter as long as it is not very deep in the atmosphere. As the
electrons have to experience mirroring to give these distributions,
they are confined in the region and reflected artificially in the
simulation. We do not invoke real magnetic mirroring because the
gyration radius is too small compared to other length scales in the
simulation and including the magnetic field would make the runs
impractically slow.

In Table \ref{tab:e}, columns 4--7 show the ratio between 511~keV line
flux and the continuum in two places: the energy flux per keV at
200~keV and the integrated continuum from 8--15~MeV.  The simulation
marked with dashes gave no photons from 8--15~MeV. The columns marked
``(sim.)'' are the ratios from the direct output of the simulations.
To get the ratios marked ``(cnvlv.),'' we convolved our simulated
spectra with the instrumental response matrix of {\it RHESSI}, so we
could compare the simulations with the ratio of counts in a flare
observed with that spacecraft.  The last three columns give the
production efficiency for photons in each band exiting the Sun per
input particle in the simulation.  The first half of Table \ref{tab:e}
represents a disk flare (cosine of the viewing angle $>0.8$) and the
last half represents a limb flare (cosine of the viewing angle between
0.2 and 0.4).

\begin{table}[htbp]
  \centering
  \begin{tabular}{cccccccccc}
    Angular&Spectral&Viewing&\multicolumn{2}{c}{511 keV Flux/ Flux}
    &\multicolumn{2}{c}{511 keV Flux/}&\multicolumn{3}{c}{Conversion Rate}\\
    Distribution&Index&Angle (cos$\beta$)&\multicolumn{2}{c}{Per keV at
      200 keV}& \multicolumn{2}{c}{8--15 MeV
      Flux}&\multicolumn{3}{c}{Photon/Proton}\\
    & & & (sim.)&(cnvlv.)&(sim.)&(cnvlv.)&511&200&8--15\\
    \hline
    Downward beamed&2.2&$>$0.8&1.13&0.84&16.5&8.67&3.8e-6&3.4e-6&2.3e-7\\
    &3.2& &0.45&0.15&14.5&5.07&2.9e-7&6.4e-7&2e-8\\
    Downward iso.&2.2& &1.92&0.94&22.7&11.5&1.1e-5&5.9e-6&5e-7\\
    &3.2&&0.45&0.18&58.0&13.1&5.8e-7&1.3e-6&1e-8\\
    Isotropic&2.2& &0.95&3.44&0.093&0.47&9.03e-6&9.5e-6&9.7e-5\\
    &2.7&&0.42&1.98&0.091&0.53&1.9e-6&4.5e-6&2.1e-5\\
    &3.2&&0.11&0.73&0.12&0.77&1.6e-7&1.5e-6&1.4e-6\\
    Pancake &2.2&&2.57&1.12&10.0&5.86&1.5e-5&6.0e-6&1.5e-6\\
    &2.7&&1.21&0.60&14.0&8.71&3.2e-6&2.7e-6&2.3e-7\\
    &3.2&&0.40&0.30&10.5&8.72&6.3e-7&1.6e-6&6e-8\\
    Downward beamed&2.2&0.4--0.2&0.46&0.43&1.38&1.36&6.2e-7&1.4e-6&4.5e-7\\
    &3.2& &0.09&0.02&-&-&6.0e-8&6.6e-7&-\\
    Downward iso.&2.2&&0.96&0.91&1.09&1.85&3.9e-6&4.1e-6&3.6e-6\\
    & 3.2&&0.15&0.29&1.60&4.89&2.4e-7&1.6e-6&1.5e-7\\
    Isotropic&2.2&&0.55&3.48&0.046&0.46&4.4e-6&8.1e-6&9.6e-5\\
    &2.7&&0.25&1.85&0.058&0.524&1.2e-6&4.9e-6&2.1e-5\\
    &3.2&&0.05&0.77&0.06&0.71&7.0e-8&1.3e-6&1.4e-6\\
    Pancake &2.2&&0.81&3.69&0.058&0.47&9.4e-6&1.2e-5&1.6e-4\\
    &2.7&&0.30&1.98&0.062&0.54&2.15e-6&7.2e-6&3.4e-5\\
    &3.2&&0.14&1.05&0.072&0.66&5.2e-7&3.8e-6&7.3e-6\\
    \hline
  \end{tabular}
  \caption{Flux ratios of photons in three energy bands from simulations of electron injection (see text). }
  \label{tab:e}
\end{table}

The electron angular distribution affects the outgoing photons in several
ways. First, the bremsstrahlung photons are highly beamed in the
direction of the electrons' motion.  The more the primary electrons are
beamed downward, the fewer photons can escape to be
observed. Second, the more the primary electrons are beamed downward, the
deeper positrons are produced and the fewer annihilation photons escape
without being scattered or absorbed.  Therefore, the escaping flux
and the annihilation line to continuum ratio depends on the
injected angular distribution. 

Not all the photons recorded come to the detector directly after they
are produced. Some of them may have been Compton scattered, changing
both their energy and direction.
Compton scattering becomes very important when the
angular distribution of injected electrons is mostly downward.  In
this case, since most direct bremsstrahlung photons head into
the Sun, Compton ``albedo'' can contribute a
significant part of the observed spectrum, and may even become
dominant at lower energies \citep{kontar06,kotoku07}.

\begin{figure}[htbp]
\centerline{\includegraphics[width=0.8\textwidth]{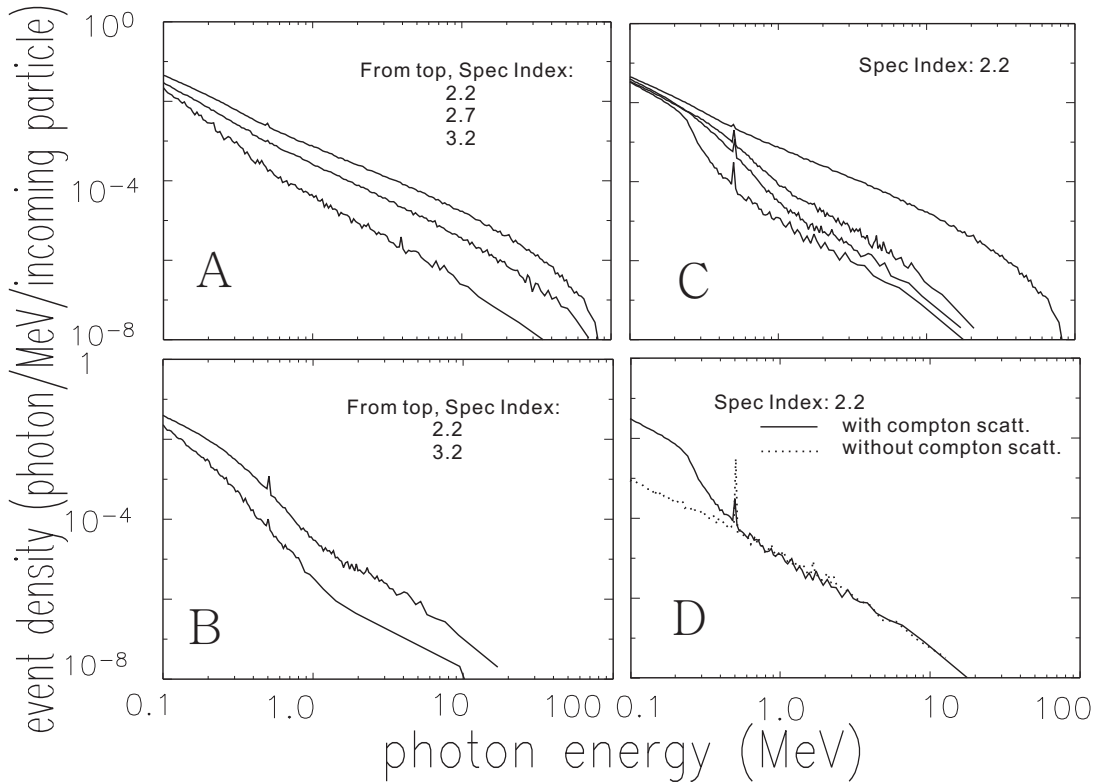}}
\caption[]{ The outgoing gamma-ray spectrum for different electron
  spectra and distributions.  \textit{Panel A}: Isotropic distribution.
  \textit{Panel B}: Directly downward-beamed distribution.  
  \textit{Panel C}: the outgoing spectrum for different
  injected angular distributions with electron spectral index
  2.2. From top down: (1) isotropic distribution; (2) pancake
  distribution uniform within angles $\theta$ such that $-0.3<\cos \theta
  <0.3$; (3) downward-isotropic distribution; and (4) downward-beamed
  distribution. \textit{Panel D}: the
  gamma-ray spectrum from directly downward-beamed electrons with a
  spectral index 2.2, with the Compton scattering process turned on
  (solid) and off (dotted). All electron spectra are cut off at
  0.1~MeV and 100~MeV and the outgoing spectra are collected at angles
  $\beta$ from the solar normal such that $\cos \beta > 0.8$.}\label{fig:e_all}
\end{figure}

In panels A and B of Figure \ref{fig:e_all}, we plot the outgoing
gamma-ray spectra for different parameters of injected electrons,
collected for $\cos \beta > 0.8$. The
spectra are normalized to outgoing photons per MeV per incoming
electron.  As expected, the harder the spectral index, the
higher the production rate, since high-energy electrons are more
efficient for thick-target bremsstrahlung.
It can also be seen that the production rate
decreases with more downward beaming (panel C of Figure
\ref{fig:e_all}). The spectra show a softening at low energies, 
the extra flux coming from Compton-scattered photons that were
originally directed downward.  This effect can be seen most clearly in
the last panel of Figure \ref{fig:e_all}, in which we show the results
with Compton scattering turned on and off.  As expected, this effect is most
significant when the electron distribution is most downward.

\begin{figure}[htbp]
\centerline{\includegraphics[width=0.8\textwidth]{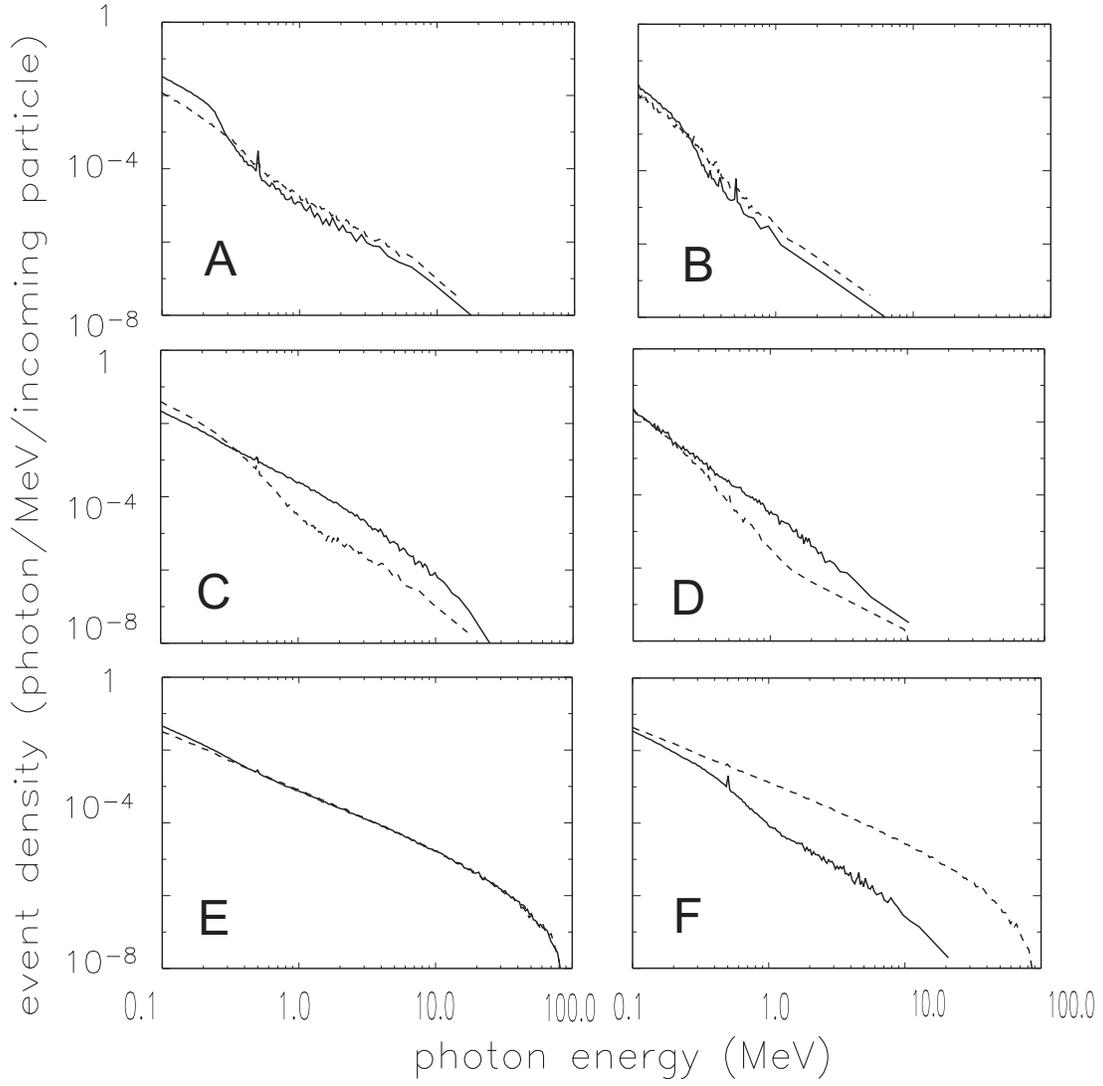}}
\caption[]{ The gamma-ray spectrum recorded at different outgoing
  angles $\beta$ from the solar normal. The solid line is $\cos \beta
  > 0.8$ and the dashed line is $0.2 < \cos \beta < 0.4$.
  \textit{Panel A}: downward-beamed distribution with spectral index
  2.2.  \textit{Panel B}: downward-beamed distribution with spectral
  index 3.2. \textit{Panel C}: downward-isotropic distribution with
  spectral index 2.2. \textit{Panel D}: downward-isotropic
  distribution with spectral index 3.2.  \textit{Panel E}: isotropic
  distribution with spectral index 2.2.  \textit{Panel F}: pancake
  distribution with spectral index 2.2 (isotropic within angle $\theta$
  of solar normal such that $-0.2<\cos \theta <0.2$).
}\label{fig:a_all}
\end{figure}

Figure \ref{fig:a_all} shows the gamma-ray spectrum recorded at
different values of the outgoing angle $\beta$ between the photon
direction and outward solar normal.  The two traces in each panel
represent a disk flare (solid) and a limb flare (dashed). 

There are two process that affect the production rate at different
outgoing angles: relativistic beaming in a downward electron
distribution will tend to put out more photons at large
$\beta$, while, on the other hand, photons coming out at large
$\beta$ have longer paths in the Sun and are more likely to scatter.
At lower energies, the second process is more important and will make
disk flares appear brighter. At high energy, however, the beaming
effect is more important and will cause limb brightening. This result
may explain the discovery of limb brightening at $>$ 0.3~MeV
\citep{vestrand87} but not over the range 5--500~keV, which
is dominated by much lower energy photons \citep{li94,li95}.
The same effect can cause a spectral break within a given flare:
our simulations show spectral hardening for limb flares in the
high-energy band (higher than about 400~keV) but not below,
in agreement with the observations of \citet{li95}.

For the simulated isotropic distribution, 
most photons observed are direct bremsstrahlung, which is also
isotropic, so there is no significant spectral evolution with
viewing angle, as shown in panel E of Figure
\ref{fig:a_all}. For pancake distributions, it is still true that
the main component observed is directly from bremsstrahlung, but
beaming to high $\beta$ is visible at all energies,
as shown in panel F of Figure \ref{fig:a_all}.

In Table \ref{tab:e}, we summarize the ratio between 511~keV line flux
and flux at 200~keV as well as the ratio between the 511~keV line and
the 8--15~MeV continuum from our simulations. In order to compare
these ratios with {\it RHESSI} data, we convolved the spectra from our
simulations with the spectral response matrix of the \textit{RHESSI}
instrument \citep{smith02}; both the direct output of the simulations
and the ratio in the ``count space'' of the instrument after
convolution are shown in Table \ref{tab:e}.  It is interesting to note
that the ratios can either decrease or increase from the convolution
process, depending on the overall shape of the spectrum.  When the
high-energy bremsstrahlung escapes the Sun easily (such as for an
isotropic electron distribution), not only does it overwhelm the solar
annihilation line but also the multi-MeV bremsstrahlung photons pair
produce in the spacecraft, so that the count spectrum has a more
prominent 511~keV line than the photon spectrum.  For flare spectra in
which little MeV bremsstrahlung escapes, on the other hand, the most
important instrumental effect is that the solar 511~keV photons (which
are significant in this case) often Compton scatter out of {\it
  RHESSI}'s detectors after a single interaction, so that they
register as continuum photons instead of line photons, causing the
count spectrum to have a less significant line than the photon
spectrum.  A solar gamma-ray spectrometer with a heavy anticoincidence
shield, such as the Gamma-Ray Spectrometer on the {\it Solar Maximum
  Mission} \citep{forrest80}, would be less susceptible to all these
instrumental effects, and the line-to-continuum ratios would be much
more similar in the count and photon spectra.

\subsubsection{Interactions of Accelerated Protons}
\label{sec:proton}

\begin{figure}[htbp]
\centerline{\includegraphics[width=0.9\textwidth]{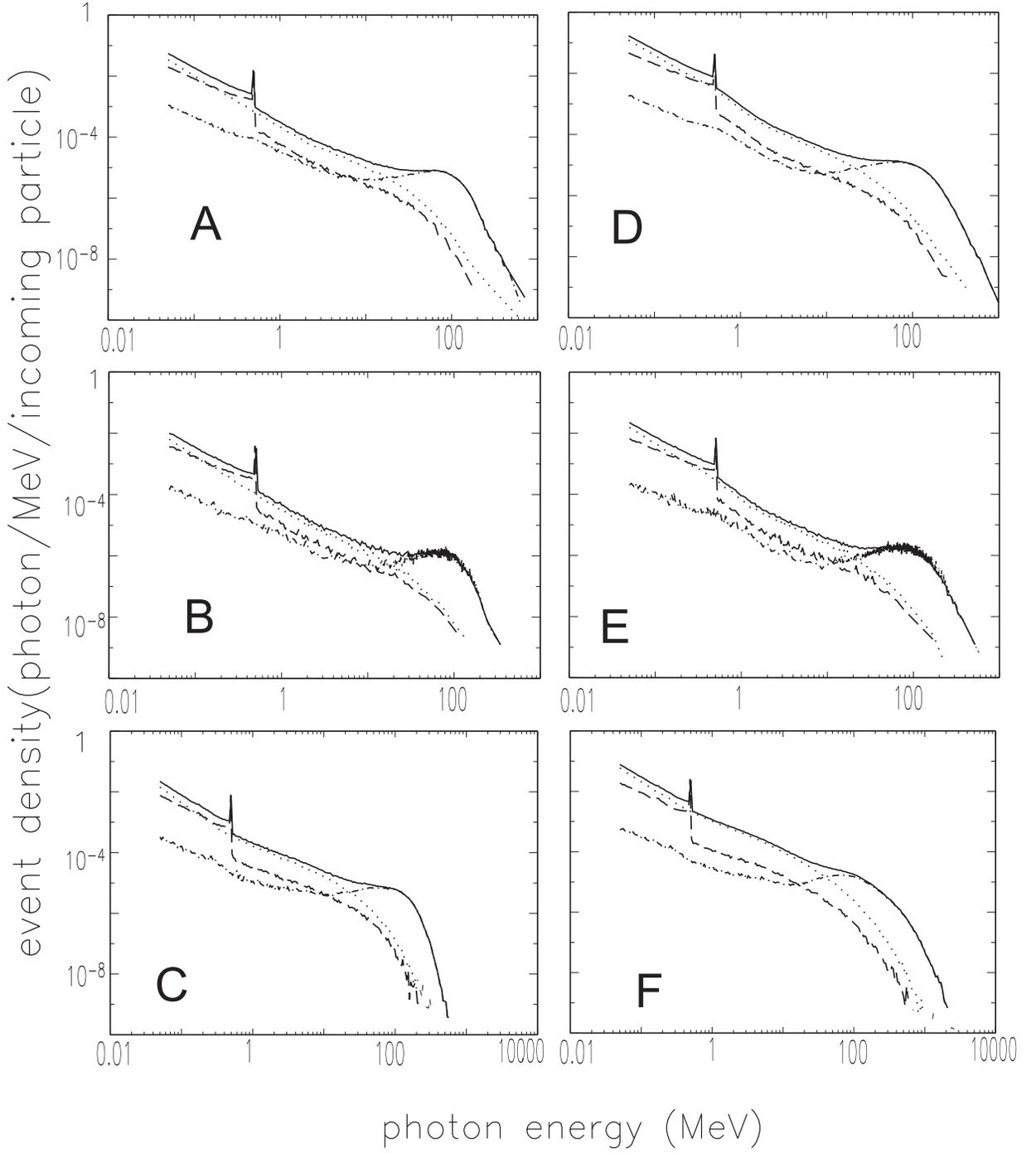}}
\caption[]{ The outgoing gamma-ray spectrum for different parameters
  of injected protons. \textit{Dash dot:} $\pi^0$ decay.
  \textit{Dotted}: bremsstrahlung of $e^{\pm}$ from the decay of
  $\pi^{\pm}$.  \textit{Dashed}: positron annihilation.
  \textit{Thin solid:} the total spectrum.  \textit{Panels A and B}:
  downward-beamed distributions with spectral index 2.2 and 3.2,
  respectively.  \textit{Panels D and E}: downward-isotropic
  distributions with spectral indices 2.2 and 3.2, respectively. The above
  four panels are for
  $\cos \beta>0.8$. \textit{Panels C and F}: downward-beamed and 
  downward-isotropic distributions with spectral index 2.2, recorded at
  $0.2<\beta<0.4$.  All proton spectra are cut off at 100~MeV
  and 10000~MeV.}\label{fig:proton_out}
\end{figure}

We also simulated the interaction between accelerated protons and the
solar atmosphere. The simulated solar atmosphere was the same as in the
electron simulations.  As discussed above, we are at this time simulating
only the production of pions and their secondaries, not nuclear
excitation, spallation, and radioactive decay.
We simulated downward beamed and
downward isotropic proton distributions, with results shown in 
Table \ref{tab:p}
and Figure \ref{fig:proton_out}. A pancake distribution of protons
is not included, since it has been ruled out for at least one
gamma-ray flare by observations of strong redshifts in the nuclear
de-excitation lines \citep{smith03}.

The shape of the outgoing spectrum changes little
when the angular and spectral distributions of the protons are varied,
but the overall gamma-ray luminosity is greater for harder
spectral indices and for the more isotropic distribution. 
This is expected, since pions are produced only by the highest-energy
protons and since the downward-isotropic distribution will produce
some pions at shallower column depths where photons are better able
to escape (see below).  The spectra observed at
$0.2<\cos\beta<0.4$ extend to higher energy because
more photons emerge without scattering.

\begin{table}
  \centering

  \begin{tabular}{cccccccccc}
    Angular&Spectral&Viewing&\multicolumn{2}{c}{511 keV Flux/ Flux}
    &\multicolumn{2}{c}{511 keV Flux/}&\multicolumn{3}{c}{Conversion Rate}\\
    Distribution&Index&Angle (cos$\beta$)&\multicolumn{2}{c}{Per keV
      at 200 keV}& \multicolumn{2}{c}{8--15 MeV
      Flux}&\multicolumn{3}{c}{Photon/Proton}\\
    & & & (sim.)&(cnvlv.)&(sim.)&(cnvlv.)&511&200&8--15\\
    \hline
    Downward iso.&2.2&$>0.8$&57.6&15.3&4.51&0.72&1.9e-3&3.36e-5&4.3e-4\\
    &3.2&&71.6&17.0&6.40&0.80&3.0e-4&4.17e-6&4.67e-5\\
    Downward beam&2.2&&61.7&17.1&3.07&0.60&6.6e-4&1.1e-5&2.1e-4\\
    &3.2&&72.7&17.9&5.22&0.71&1.4e-4&2.0e-6&2.7e-5\\
    Pancake&2.2&&57.3&14.9&6.83&0.86&3.5e-3&6.14e-5&5.15e-4\\
    &3.2&&73.3&17.3&10.0&0.93&4.7e-4&6.4e-6&4.7e-5\\
    Downward iso.&2.2&0.4--0.2&54.5&17.7&0.64&0.45&5.77e-4&1.03e-5&8.8e-4\\
    &3.2&&74.7&18.3&1.06&0.49&9.28e-5&1.24e-6&8.73e-5\\
    Downward beam&2.2&&65.5&19.0&0.80&0.45&1.71e-4&2.61e-6&2.13e-4\\
    &3.2&&75.1&19.1&1.33&0.49&3.99e-5&5.32e-7&3.01e-5\\
    Pancake&2.2&&54.7&16.9&0.70&0.46&1.1e-3&2.07e-5&1.6e-3\\
    &3.2&&76.0&17.8&1.21&0.51&1.6e-4&2.17e-6&1.36e-4\\

\hline
  \end{tabular}
  \caption{Flux ratios from simulations of proton injection.}
  \label{tab:p}
\end{table}

\begin{figure}[htbp]
\centerline{\includegraphics[width=0.8\textwidth]{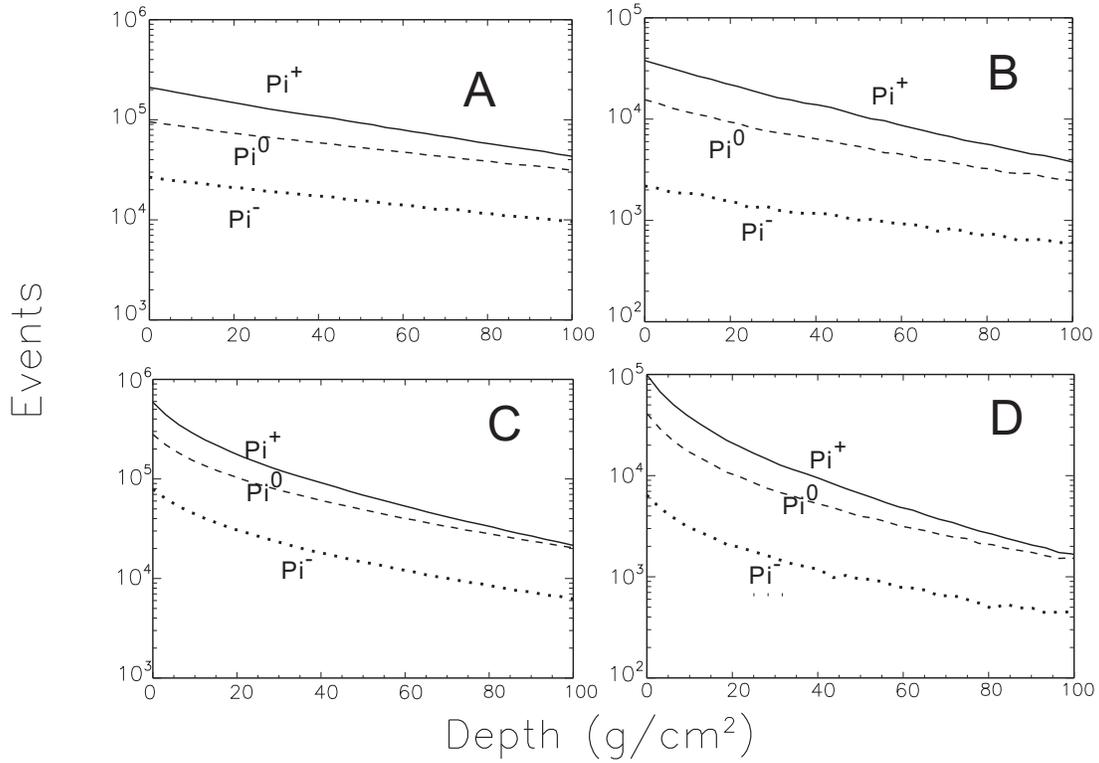}}
\caption[]{ The pion-producing depth distribution for different
  parameters of injected protons. \textit{Panels A and B}: downward-beamed 
  distributions with spectral indices 2.2 and 3.2, respectively.
  \textit{Panels C and D}: downward-isotropic distributions with
  spectral indices 2.2 and 3.2, respectively. In each panel,
  \textit{solid}: $\pi^{+}$; \textit{dashed}: $\pi^0$;
  \textit{dotted}: $\pi^{-}$. }\label{fig:piproduct}
\end{figure}

Pions are produced very deep in the Sun, at column depths of tens of
g/cm$^2$, as shown in Figure \ref{fig:piproduct}.  
The production rate
falls off more quickly with depth for the downward isotropic distribution,
since protons at a shallow angle can go through a large column of
solar atmosphere and still produce pions at relatively small depth.

\begin{figure}[htbp]
\centerline{\includegraphics[width=0.6\textwidth]{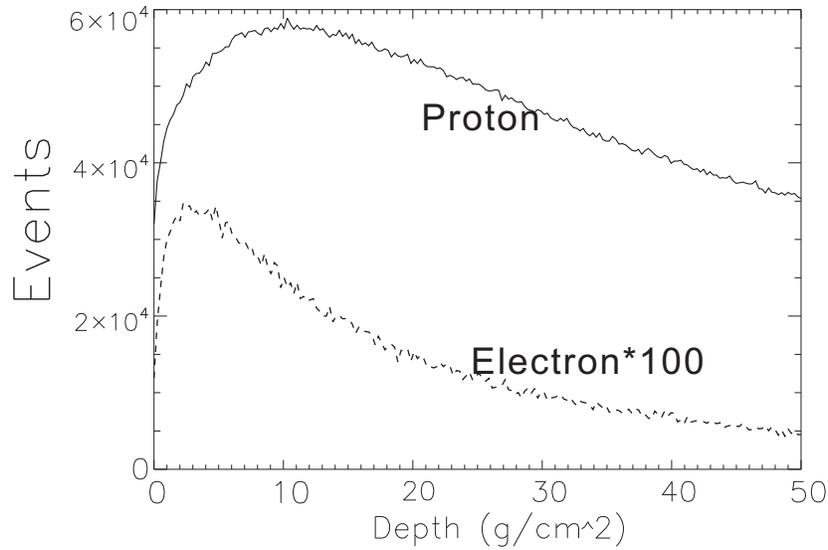}}
\caption[]{ The distribution of the depth where positron annihilation occurs.
  \textit{Solid line}: positrons generated from injected protons.
  \textit{Dashed line}: positrons generated from injected electrons. 
  Both the electrons and
  protons follow a downward isotropic distribution with spectral index
  2.2. }\label{fig:annidepth}
\end{figure}

Pions have very short lifetimes (2.6$\times 10^{-8}$ s for $\pi^{\pm}$
and 8.4$\times 10^{-17}$ s for $\pi^0$), so they decay where they are
produced. The positrons produced by $\pi^{+}$ decay have to travel
some distance before they slow down and annihilate. Figure
\ref{fig:annidepth} shows the depth distributions of positron
annihilation events for both injected protons and electrons when both
have a spectral index of 2.2 and energy ranges of 100--10000~MeV and
0.1--100~MeV respectively.  Each proton in this case is more than 200
times as likely to produce a positron as an electron, and they tend to
be produced deeper in the solar atmosphere.  Among positrons
originating with protons, we found that 70\% originate from $\pi^{+}$
decay, 20\% from pair-production of the gamma-rays produced in $\pi^0$
decay, and 10\% from more indirect cascade processes (for example,
$\pi^{-} \rightarrow e^{-} \rightarrow {\rm bremsstrahlung}
\rightarrow {\rm pair~production}$).

\begin{figure}[htbp]
\centerline{\includegraphics[width=0.8\textwidth]{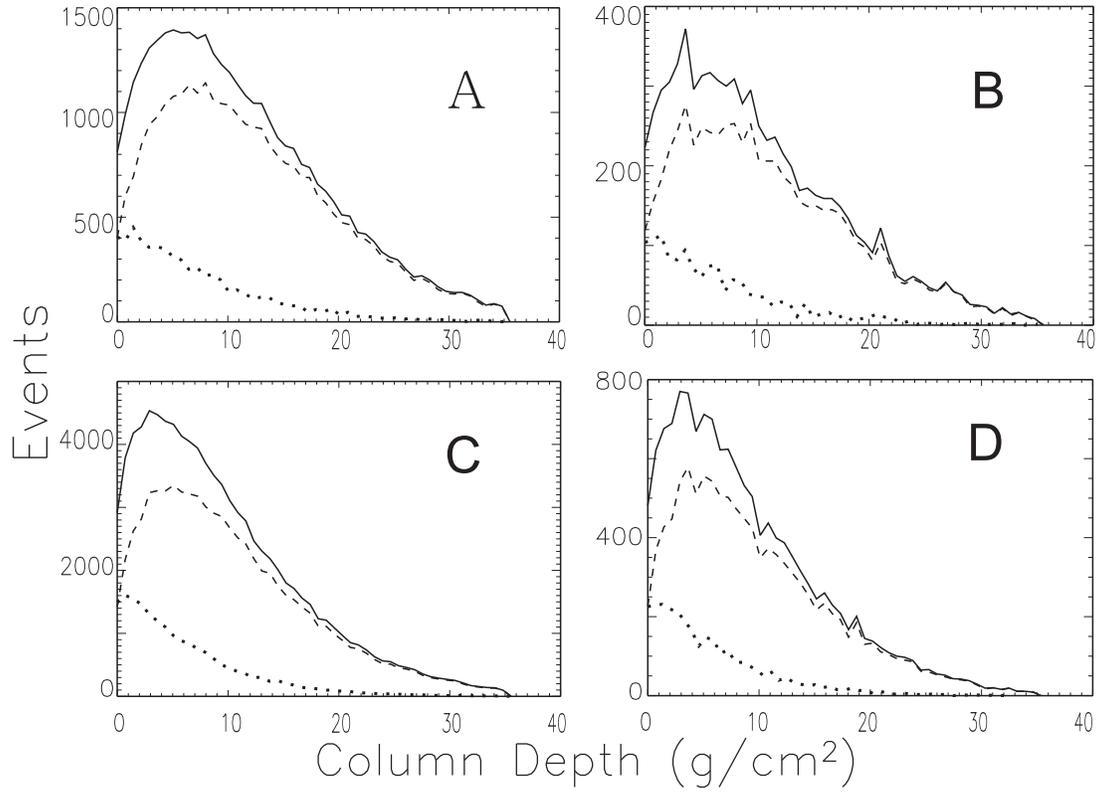}}
\caption[]{ The distribution of the depth where positron annihilation
  occurs, corresponding to photons collected at $\cos
  \beta>0.8$. \textit{Dashed line}: photons that experience Compton
  scattering before being detected.  \textit{Dotted line}: photons
  that do not experience Compton scattering.  \textit{Solid line}:
  total. \textit{Panels A and B}: downward beamed distributions with
  spectral indices 2.2 and 3.2, respectively. \textit{Panels C and D}:
  downward isotropic distributions with spectral indices 2.2 and 3.2,
  respectively.}\label{fig:outannidepth}
\end{figure}

Most of the annihilation photons that leave the Sun in the simulations
experienced Compton scattering, so that they are observed in a
continuum below the line.  In Figure \ref{fig:outannidepth}, we plot
the depth distributions of only those annihilation events
corresponding to photons that escape the Sun.  As expected, the deeper
the annihilation happens, the fewer photons escape without scattering.
More than 70\% of the gamma-ray photons we observed experienced Compton
scattering.  The resulting Compton continuum can mimic two other
spectral components just below 511~keV: the continuum from the
3$\gamma$ decay of orthopositronium and the broad lines from
$\alpha-\alpha$ reactions.  Orthopositronium will only survive
collisional disruption before decay at low densities; thus, a 511~keV
line with little continuum directly below it can be taken as a sign of
annihilation at moderately high density but not great depth.
\citet{share04} found that if the 511~keV line was
produced under 5--7~g cm$^-2$, the Compton continuum would be similar
to what is seen below the line by {\it RHESSI} from the X17 flare of
2003 October 28. Our simulations show that a significant fraction of
positrons annihilate at just this depth if they originate from pions.
Positrons from $\beta +$ decay of spallation products will probably
have a shallower distribution since they can be created by more
numerous, lower-energy protons of tens of MeV that do not penetrate
to these depths.

\subsection{The X17 Flare of 2003 October 28}
\label{sec:oct28}

\begin{figure}[htbp]
\centerline{\includegraphics[width=0.8\textwidth]{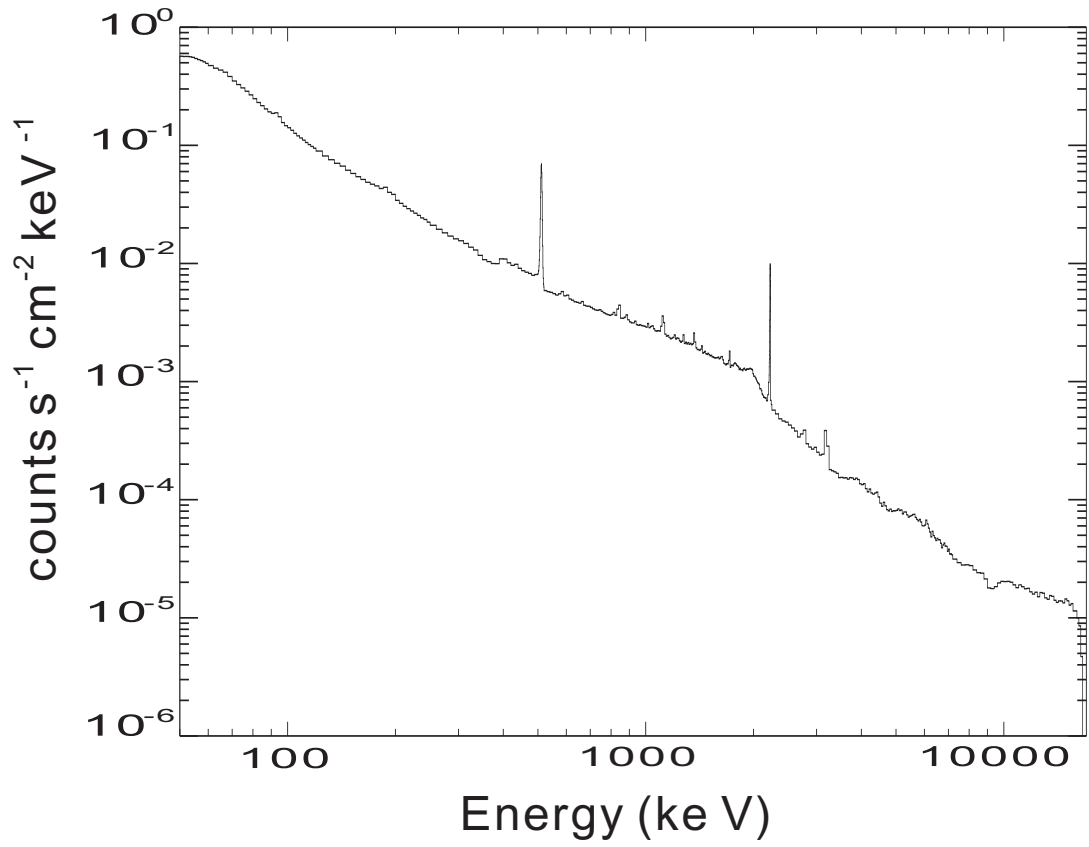}}
\caption[]{ The overall spectrum of the flare of 28 October 2003, from
  \textit{RHESSI} (11:06 to 11:26 UT).  A small discontinuity at
  375~keV, a peak at 3~MeV, and a dip around 9~MeV are known
  instrumental artifacts that are accounted for when model spectra are
  fitted to these data.  The effect of Compton scattering of solar
  photons in the instrument is most clearly seen at 2~MeV, where there
  is a shoulder due to Compton backscattering of 2.2~MeV neutron
  capture photons in {\it RHESSI}'s germanium
  detectors.}\label{fig:spec}
\end{figure}

This flare, the second largest in the ``Halloween storms'' of 2003,
occurred near disk center at a heliocentric angle of
$\arccos(0.87)$. {\it RHESSI} missed the rapid rise phase and peak of
the flare because it was crossing the South Atlantic Anomaly and only
provides data after 11:06~UT.  The \textit{RHESSI} data extend from
3~keV to 17~MeV, with energy resolution of $\sim$1--10~keV across this
range.  The high-energy spectrum of the flare is shown in Figure
\ref{fig:spec}, using data from the rear segments of the {\it RHESSI}
germanium detectors \citep{smith02}.  The 511~keV line and the 2.2~MeV
line from neutron capture on ambient protons are most clearly visible.
Since this spectrum is uncorrected for instrument response, many of
the counts are shifted to lower energies by Compton scattering in the
instrument.  A spectral accumulation taken 15 \textit{RHESSI} orbits
(about one day) earlier has been subtracted as background, since the
geographical position and radiation history of the spacecraft was
similar at that time.

\begin{figure}[htbp]
\centerline{\includegraphics[width=0.8\textwidth]{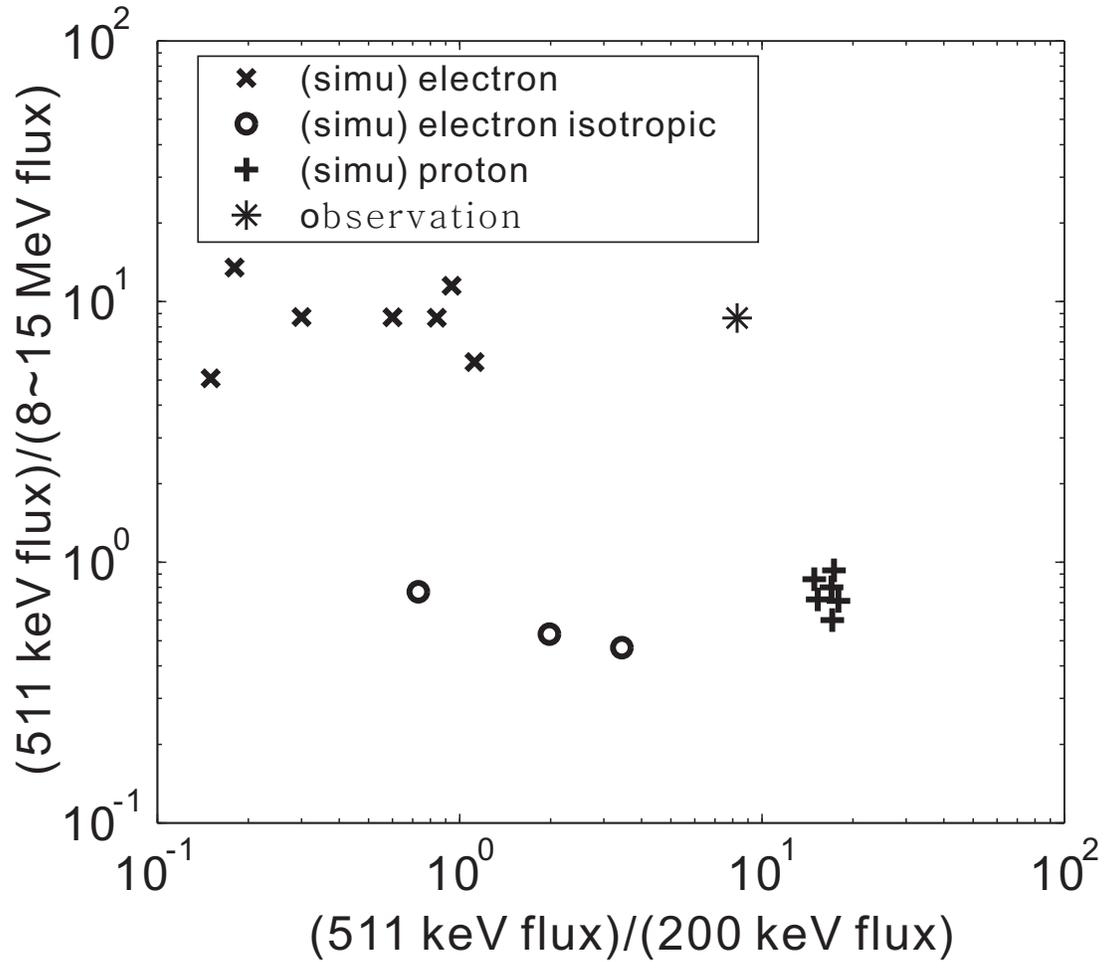}}
\caption[]{The ratio between 511~keV line flux and the continua around
  200~keV and 8--15~MeV, The meaning of different symbols are shown in
  the figure. From which, it is clear that if all the 511s are from
  proton and isotropically injected electrons, the 511/8--15 ratio
  would be well below observation and a combination of proton
  injection (plus symbol) and electron injection that is not complete
  isotropic (cross symbol) may produce the observation. See the text
  for more details.}\label{fig:ratio}
\end{figure}

In Table \ref{tab:ob}, we list the ratio between 511~keV line flux and
the continua around 200~keV and 8--15~MeV for comparison to our
simulations; the results are shown graphically in Figure
\ref{fig:ratio}. We find that if all or most of the 511~keV line
flux resulted from pion interactions, there would have been more
8--15~MeV continuum observed, regardless of the angular distribution
of the injected protons (Table \ref{tab:p}).  Most of the positrons
are therefore from other sources, either $\beta+$ decays (not
simulated here) or bremsstrahlung gammas from high-energy flare
electrons (Table \ref{tab:e}).  If all or most of the 511~keV line
flux came from accelerated electrons, the 8--15~MeV continuum would
also be overproduced for an isotropic distribution.  Either a
domination of the positron source by $\beta+$ decay or by mostly
downward electrons is allowed.

In the comparison above, we did not consider the effect of 3$\gamma$
annihilation on the 511~keV line.  Most of the annihilation in our
model occurs below the photosphere, where the density is high enough
that orthopositronium will be destroyed by collisions before decay,
thus we believe that this is a good approximation.
However, if there were any  3$\gamma$
annihilation, it would make the 511~keV line to 8--15 MeV ratios even
smaller, thus strengthening the conclusion that pion decay does not
dominate positron production in this flare.

\begin{table}[!htb]
  \centering
\begin{tabular}{cccccc}
  &  511 keV  &  Continuum  &Continuum &511 keV Flux/&511 keV Flux/\\
  & Line &8--15 MeV&at 200 keV per keV & flux Per keV at 200 keV&8--15 MeV flux\\
  \hline

  &43561&5031&5120  & 8.5 & 8.65\\
  \hline
\end{tabular}
\caption{Flux ratios for the flare of 2003 October 28 from
  \textit{RHESSI} data, for comparison with Tables \ref{tab:e} and
  \ref{tab:p}.  Data in the first two columns are in  
  raw background-subtracted counts.}
\label{tab:ob}
\end{table}

\section{Discussion and Summary}
\label{sec:dis}

We have used the GEANT4 package to simulate the spectra produced
by the interactions of high-energy flare particles in the Sun,
emphasizing electron bremsstrahlung and pion production by protons.

We find that the angular distribution of primary electrons accelerated
in solar flares can greatly affect the shape and production rate of
outgoing gamma-ray photons. In general, the more the injection is
downward beamed, the steeper the outgoing spectrum is and the lower
the production rate. \citet{kotoku07} found the same result and
suggested that limb flares should therefore have harder spectra than
disk flares.  Extending those results down to lower energies, we find
that a downward-biased distribution will cause limb brightening at
higher ($\gtrsim$ 0.3 MeV) energy and limb darkening at lower
energy. A isotropic distribution will show no bias and a pancake
distribution will produce limb brightening at all energies. These
results are due to the combined effects of bremsstrahlung and
Compton scattering.

We modeled two classes of mechanism that can produce positrons in
flares -- the electromagnetic cascade from accelerated electrons and
the decay products of pions.  The third source, and perhaps the most
important, is radioactive decay, which we postpone until we can
further study and perhaps improve the nuclear cross-sections available
in GEANT4.  We found that the annihilation line resulting from
accelerated electrons has only a weak dependence on the angular
distribution of the electrons.  Since the bremsstrahlung continuum has
a strong dependence, the ratio of the annihilation line to the
continuum can constrain the electron angular distribution if the
electron contribution to the positron population can be isolated.
\citet{murphy84} and \citet{gan04} used the time history of the C/O
de-excitation lines from 4~MeV to 7~MeV to estimate the positron
production due to decay of spallation products (i.e., due to
accelerated protons below the pion production threshold).  Such a
technique, combined with observations of photons up to 100~MeV
\citep[e.g.]{arkhangelskaja09} to fix the pion contribution to the
annihilation line, could allow the electron contribution to be
isolated if the three components (electrons, lower-energy protons, and
high-energy protons) have different time profiles.  If the electron
contribution to the annihilation line can be isolated, it becomes a
new and valuable diagnostic for the electron spectrum and angular
distribution.  Future space missions with $\sim$ 10'' imaging in the
gamma-ray range could allow spatial as well as spectral and temporal
information to be used to isolate the electron contribution to the
annihilation line.

Comparing our simulations to {\it RHESSI} data from the 2003 October
28 flare, we find that the high ratio of the 511~keV line to the
8--15~MeV continuum implies that either radioactivity, bremsstrahlung
from downward-biased electrons, or a combination dominates the
positron production in this flare.

We also find that positrons from $\pi^{+}$ decay annihilate at a
column depth of $\sim$ 10 g cm$^{-2}$ in the Sun and most of the
gamma-ray photons they produce experience Compton scattering before
escaping, producing a continuum that resembles the
3$\gamma$ decay of orthopositronium.

In gamma-ray flares, a hardening break around 0.6~MeV is often found for the
power-law component of the spectra. This break may be interpreted as
indicating two populations of electrons \citep{krucker08},
or as electron-electron bremsstrahlung becoming dominant above that
energy \citep{kontar07}. Based on our simulation, it is also possible
that the component above 0.6~MeV is not caused by primary electrons that
are accelerated in the flares but by electrons and positrons
from the decay of pions generated by the accelerated protons.  To test
this possibility, we will need to compare our simulations with flare data
up to $\sim 100$~MeV to fix the normalization of the pion component and
see if it is enough to contribute what is usually thought of as the
hard tail of the bremsstrahlung spectrum.

\acknowledgments

The authors thank Ronald Murphy, Gerald Share, Albert Shih,
Troy Porter, and Eduard Kontar for contributing by explanation and
example to this work and our understanding.  This work was supported
by NASA grant NNG05G189G-004, NASA contract NAS5-98033, and China
Scholarship Council Postgraduate Scholarship Program.

\end{document}